\documentclass[article,sigplan,10pt]{acmart}
\usepackage{graphicx}
\usepackage{amsmath,amsfonts}
\usepackage{fdsymbol}
\usepackage{multicol}
\usepackage{multirow}
\usepackage{colortbl}
\usepackage{lipsum}
\usepackage{graphicx}
\usepackage{textcomp}
\usepackage{listings}
\usepackage{cellspace}
\usepackage{fixltx2e}
\usepackage{hyperref}
\usepackage{chngcntr}
\counterwithin*{equation}{section}
\AtBeginDocument{%
  }

\setcopyright{acmcopyright}
\copyrightyear{2018}
\acmYear{2018}
\acmDOI{XXXXXXX.XXXXXXX}

\acmConference[Conference acronym 'XX]{Make sure to enter the correct
  conference title from your rights confirmation emai}{June 03--05,
  2018}{Woodstock, NY}
\acmPrice{15.00}
\acmISBN{978-1-4503-XXXX-X/18/06}

\begin{document}

\title{L4 Pointer: An efficient pointer extension for spatial memory safety support without hardware extension}


\author{Seong-Kyun Mok}
\authornotemark[1]
\email{mok7764@cnu.ac.kr}
\affiliation{%
  \institution{Chungnam National University}
  \city{Daejeon}
  \country{USA}
}

\author{Eun-Sun Cho}
\affiliation{%
  \institution{Chungnam National University}
  \city{Daejeon}
  \country{USA}
}
\email{eschough@cnu.ac.kr}


\begin{abstract}
Since buffer overflow has long been a frequently occurring, high-risk vulnerability, various methods have been developed to support spatial memory safety and prevent buffer overflow. However, every proposed method, although effective in part, has its limitations. Due to expensive bound-checking or large memory intaking for metadata, the software-only support for spatial memory safety inherently entails runtime overhead. 
Contrastingly, hardware-assisted methods are not available without specific hardware assistants. 
To mitigate such limitations, Herein we propose L4 Pointer, which is a 128-bit pointer extended from normal 64-bit virtual addresses. By using the extra bits and widespread SIMD operations, L4 Pointer shows less slow-down and higher performance without hardware extension than existing methods. 
\end{abstract}



\maketitle

\section{Introduction}
Buffer overflows have long been the most dangerous vulnerability and still rank among the most dangerous software weaknesses in the CWE Top 25 \cite{CWE, CWETOP25}. 
Attackers exploit buffer overflows and gain control of the flow of execution or security leaks. Buffer overflows even occur in widely used open sources. 
For instance, the Heartbleed bug is the most famous vulnerability categorized buffer overflow found in OpenSSL \cite{heartbleed, openssl}. 
In addition, as the use of IoT devices increases, so does the frequency of buffer overflow \cite{IoTBuffer}, because Due to the characteristic of embedded devices, C/C++ has to be used and thus buffer overflow occurs more frequently. 

Bounds checking is one of the oldest and most common defenses used to prevent buffer overflow \cite{Baggybounds, CCured, CHERI, FRAMER, Hardbound,  memorysafety, Delta, Sanitizer, In-Fat, intelMPX, ALEXIA, SHAKTI, Backwards, ARM}. By saving the bounds information of a memory object with each pointer, defenses can insert runtime bounds checking to verify that the pointer still points to the valid range. Since some programming languages like Java support bounds checking, it can be performed without a special process. However, other programming languages like C/C++ do not support bounds checking. As such, supporting bounds checking in C/C++ has been extensively studied. Since bounds checking is inserted into the program, it incurs runtime and memory overhead. 

Bounds checking is an appropriate method of decreasing overhead. Bounds information transforms and stores forms called metadata. First, for this to successful occur, the metadata must be the appropriate size. If the metadata is too small, it will not include bounds information. Some researches support small metadata, so supports only upper bounds \cite{Delta}, or has false negative \cite{FRAMER}. On the contrary, too big a size of metadata incurs memory overhead \cite{CHERI}. 
Second, it is necessary to design the manipulation method of the metadata properly. 
With bounds checking, we should consider runtime overhead of a process.
In addition, the manipulation of a pointer with metadata should be atomic, otherwise, it will lead to false positives or negatives in multi-threading programs. 
While the Intel MPX has excellent performance, it is deprecated because it does not provide multi-threaded safety \cite{intelMPX, DisccusionOfMPX}. 

Recent works have proposed `fat-pointers' enriching pointers with metadata for bounds checking.
However, the fat pointers implemented by a software-only method suffer from high runtime overhead when applied to real-world cases \cite{FRAMER, CCured}. 
To get over this, hardware-assisted solutions have been suggested against buffer overflow to lower runtime overhead.
Unfortunately, traditional hardware-assisted fat pointers do not support multi-thread safety in most cases, not to get engaged in conflicts with other functionalities from the hardware. 
Recently, advanced hardware-assisted fat pointers have been implemented on special hardware.
Although this approach achieves low runtime overheads and multithread safety, 
such fat pointers are not easily available without specific hardware assistants. 

This paper proposes the L4 Pointer, which incurs low overhead and atomicity without need for special hardware.
Based on 128-bit pointers, the L4 Pointer reserves sufficient space to store metadata as well as the pointer itself. 
Depending on the vector operations and 128-bit registers supported by most hardware architectures \cite{Intel, ARM}, the L4 Pointer is anticipated to be preferred in real-world usage. 
The contributions of this paper include the following. 

\begin{itemize}
\item This paper presents the design and implementation of a prototype of the L4 Pointer, 
which provides sufficient metadata storage space to support both upper and lower-bound checks against buffer overflow. 
\item The proposed L4 Pointer provides atomic operations of the pointer and metadata. 
\item The proposed L4 Pointer works on commonly used architectures such as Intel processors and ARM processors. 
The target architecture of the prototype of the L4 Pointer is Intel x86 \cite{ARM, Intel}.
\end{itemize}

\begin{table*}[]
\caption{Comparisons of approaches}
\label{approaches}
\resizebox{1.5\columnwidth}{!}{%
\begin{tabular}{|cc|c|c|c|}
\hline
\multicolumn{1}{|c|}{}   &              & Approaches                                                            & Runtime overhead & Multithread-safety \\ \hline
\multicolumn{1}{|c|}{\multirow{2}{*}{\begin{tabular}[c]{@{}c@{}}Tracking \\ metadata\end{tabular}}} &
  Per-object &
  \begin{tabular}[c]{@{}c@{}}Each pointer has each \\ metadata\end{tabular} &
  High &
  varying \\ \cline{2-5} 
\multicolumn{1}{|c|}{}   & Per-pointer  & \begin{tabular}[c]{@{}c@{}}Memory object has \\ metadata\end{tabular} & Low              & O                  \\ \hline
\multicolumn{1}{|c|}{\multirow{2}{*}{\begin{tabular}[c]{@{}c@{}}Size of \\ metadata\end{tabular}}} &
  Fat pointer &
  Extending the size of the pointer &
  High &
  X \\ \cline{2-5} 
\multicolumn{1}{|c|}{} &
  \begin{tabular}[c]{@{}c@{}}Tagged \\ pointer\end{tabular} &
  \begin{tabular}[c]{@{}c@{}}Saving metadata in \\ unused bits\end{tabular} &
  Low &
  O \\ \hline
\multicolumn{2}{|c|}{Hardware-assisted} & Machine dependent approaches                                          & Low              & O                  \\ \hline
\end{tabular}%
}
\end{table*}

\section{Related Works}

To prevent buffer overflow, tracking and saving metadata should be implemented. This paper groups, according to the method of metadata manipulation, (A) per-pointer and (B) per-object, (C) fat pointer, and (D) tagged pointer. In addition, there are also (E) hardware-assisted approaches. Further details are provided in the following subsections, and related works are summarized in Table \ref{comparisonsofrelatedworks}. 
As shown in Table \ref{approaches}, some approaches mixed the way to compensate for limitations. 
For example, ALEXIA and Shakti-MS mixed the per-object and per-pointer methods for the bounds check of the sub-object \cite{ALEXIA, SHAKTI}. 
Further details are provided in later sections.

\subsection{Per-pointer}
The per-pointer method involves metadata per pointer \cite{Delta, CCured, CHERI}. Even if different pointers point to the same memory object, pointers with different memory addresses have different metadata.
For example, there are variables of pointer type, namely p and q, pointing to the same memory object but not the same location. Thus, the metadata of variables p and q are different in the per-pointer method. In this method, whenever pointer arithmetic, the metadata must also be calculated, so precision is ensured compared to per-object, but compatibility is lowered than per-object. The per-pointer method can prevent sub-objects overflow. Assume the following structure in the program written in C as shown below. When the structure below is allocated in memory, the buffer check for the memory of the object as well as the boundary check for the sub-object “buf” in the object are required. This method is capable of performing boundary checks on sub-objects. 
In addition, the runtime overhead is high \cite{CCured}. To solve limitations, CCured works to determine a safe pointer to reduce overhead \cite{CCured}. A safe pointer means that there is no possibility of a buffer overflow.

\subsection{Per-object}
The per-object method has bounds metadata for each memory object \cite{FRAMER, ARM, In-Fat, EffectiveSan}. That is, if different pointers point to the same memory object, they have the same metadata. For example, the situation is the same as above subsection 2.1, but the metadata of p and q are the same. The per-object method does not perform boundary checking on sub-objects. Since it is determined whether the memory object is accessed or not, it is impossible to check the internal object boundary. Other than that, for performance reasons, some studies allocate larger than the original allocated memory size and as such, the precision is lowered. This is a method that sacrifices precision for some runtime overhead. Unlike the per-pointer method, the metadata policy, here, is not essential. Therefore, while this method has less runtime overhead and compatibility compared to the per-pointer method, it is less precise.

\subsection{Fat pointer}
The fat pointer method utilizes a larger pointer than the 64-bit general pointer to store metadata \cite{Hardbound, Baggybounds}. This method can be expressed as the code below with C syntax. 

\begin{verbatim}
struct FatPointer{
  void* metadata;
  void* pointer;
};
\end{verbatim}

In traditional fat pointers, metadata is saved in a bound table. In the above code, metadata is not stored directly in the pointer, but in the address where the metadata is stored.  As this method cannot guarantee thread safety, researchers have begun investigating a method of implementing hardware to compensate \cite{In-Fat, ALEXIA, SHAKTI, CHERI}. However, some studies are not applicable to existing processors because they are implemented as processor emulators. Since the size of the pointer is different from that of the general pointer, compatibility problems and memory overhead occur.

\subsection{Tagged pointer}
Tagged pointer is a method of storing metadata in unused bits. While the general pointer size is 64 bits, not all 64 bits are used due to physical limitations in addressable memory. The number of bits used is different for each architecture in addition to different operating systems. The advantage of tagged pointers is that the size of the pointer does not change, so the compatibility is high and the calling conventions do not need to be changed. However, some approaches are implemented in such a way as to forgo precision \cite{FRAMER} or compatibility because the size of storable metadata is restricted due to the limitation of the size of bits \cite{Delta}. Delta pointer uses the lower 32 bits as a virtual address, and the upper 32 bits to store metadata. Because only the 32-bit is used as a virtual address, compatibility is poor. Further, as 32-bit only stores information about upper bounds when used as metadata, it so it cannot detect underflow. In addition, FRAMER stores the frame of the memory in the upper 16 bits. Even if a buffer overflow occurs in a ’frame’ written in the tag, FRAMER cannot detect buffer overflow \cite{FRAMER, Delta}.

\subsection{Hardware-assisted}
There is also a way to solve to prevent buffer overflow using hardware. Intel MPX and ARM MTE provide memory safety by providing an extended instruction set \cite{intelMPX, ARM}. Although it is possible to provide faster performance than the existing methods with an extended instruction set, these methods also have disadvantages. Intel MPX is a per-pointer bounds-tracking method, but pointer arithmetic operations and metadata operations are not performed as atomic operations. Additionally, it is no longer supported by the compiler because it is not superior to other methods in terms of speed. ARM MTE tags each memory area, records the tag of the pointer, and checks memory safety by matching the tag. Because it provides instruction set extensions, it achieves a high speed. In addition, the architecture is extended to implement the fat pointer mentioned above. Recent studies have implemented a fat pointer with extended hardware \cite{In-Fat, SHAKTI, ALEXIA}. The tool in the existing architecture cannot guarantee speed and sufficient metadata storage space. However, this method cannot be applied to the existing architecture such as Intel and ARM. 

\subsection{Limitations of previous works}
Table~\ref{comparisonsofrelatedworks} presents a comparison of related works and their limitations. The operation method of the L4 pointer is derived from the delta pointer. However, the delta pointer does not cover buffer underflow, which often occurs \cite{underflow1, underflow2}. Some algorithms use a decrement operation to access an array, such as the string-matching algorithm, underflow must also be covered[8]. Intel MPX does not ensure multi-threaded safety. Slow-down only increases 1.5 times, but both false positives and negatives may occur. While FRAMER ensures multithread safety, false negatives may still occur, and overhead increases 3.23 times. The hardware extension methods typically used to overcome these shortcomings cannot be applied to the existing architecture. L4 Pointer is available on the used architectures supported by SIMD, detects buffer overflows and underflows, and supports multithread safety. Also, L4 Pointer incurs lower runtime overhead than software-only approaches.

\begin{table*}[]
\caption{Comparisons of Related Works}
\resizebox{2\columnwidth}{!}{%
\begin{tabular}{|l|c|c|c|c|c|}
\hline
                                          & \cellcolor[HTML]{C0C0C0}\begin{tabular}[c]{@{}c@{}}Subject of tracking \\ metadata\end{tabular} & \cellcolor[HTML]{C0C0C0}\begin{tabular}[c]{@{}c@{}}Subject of saving \\ metadata\end{tabular} & \cellcolor[HTML]{C0C0C0}Compatibility & \cellcolor[HTML]{C0C0C0}\begin{tabular}[c]{@{}c@{}}Runtime \\ overhead\end{tabular} & \cellcolor[HTML]{C0C0C0}Note                                      \\ \hline
\cellcolor[HTML]{C0C0C0}Intel MPX         & Per-pointer                                                                                     & Bounds table                                                                                  & Intel                                 & 1.5x                                                                                & \begin{tabular}[c]{@{}c@{}}Mutithread\\ safety X\end{tabular}     \\ \hline
\cellcolor[HTML]{C0C0C0}ARM MTE           & Per-object                                                                                      & Tagged pointer + shadow memory                                                                & ARM                                   & -                                                                                   &                                                                   \\ \hline
\cellcolor[HTML]{C0C0C0}Delta pointer     & Per-pointer                                                                                     & Tagged pointer                                                                                & All architectures                     & 1.35x                                                                               & \begin{tabular}[c]{@{}c@{}}Only upper\\ bounds check\end{tabular} \\ \hline
\cellcolor[HTML]{C0C0C0}FRAMER            & Per-object                                                                                      & Tagged pointer                                                                                & All architectures                     & 3.23x                                                                               &                                                                   \\ \hline
\cellcolor[HTML]{C0C0C0}CHERI             & Per-pointer                                                                                     & Fat pointer                                                                                   & The proposed architectures            & -                                                                                   &                                                                   \\ \hline
\cellcolor[HTML]{C0C0C0}Address sanitizer & Memory                                                                                          & Shadow memory                                                                                 & All architectures                     & 2x higher                                                                           &                                                                   \\ \hline
\cellcolor[HTML]{C0C0C0}In-Fat pointer    & Per-object                                                                                      & Fat pointer                                                                                   & The proposed architectures            & 1.24x                                                                               &                                                                   \\ \hline
\cellcolor[HTML]{C0C0C0}ALEXIA            & Per-object+per-pointer                                                                          & Tagged pointer + Fat object                                                                   & The proposed architectures            & 1.14x                                                                               &                                                                   \\ \hline
\cellcolor[HTML]{C0C0C0}Shakti-MS         & Per-object+per-pointer                                                                          & Fat pointer                                                                                   & The proposed architectures            & 1.13x                                                                               &                                                                   \\ \hline
\cellcolor[HTML]{C0C0C0}EffectiveSan      & Per-object                                                                                      & Shadow memory                                                                                 & All architectures                     & 2.15x                                                                               & Only subobject                                                    \\ \hline
\cellcolor[HTML]{C0C0C0}L4 Pointer        & Per-pointer                                                                                     & Fat pointer                                                                                   & The architectures supporting SIMD     & 1.44x                                                                               &                                                                   \\ \hline
\end{tabular}%
}
\label{comparisonsofrelatedworks}
\end{table*}

\section{L4 Pointer}
To use the L4 Pointer, a normal pointer of 64 bits is expanded to a consecutive 128 bits. Metadata is stored in the upper 64 bits, which are evenly shared by the upper and low bounds. It is significantly similar to delta pointers \cite{Delta}, except that the metadata of a delta pointer is 32 bits long, which is too short to hold both the upper and lower bounds. To handle 128-bit long pointers and manipulate metadata and pointers atomically, the L4 pointer uses vector operations and vector registers \cite{Intel, ARM}. Fortunately, the most commonly used architectures such as X86 and ARM provide these facilities. More details are provided below.

\subsection{SIMD (Single Instruction Multiple Data)}
SIMD (Single Instruction Multiple Data) is designed for parallel processing and is a method of computing multiple data with one instruction. Intel and ARM architectures, which are commonly used, support this operation as well as special registers and instructions set extensions. The method proposed in this study needs to update metadata and pointers simultaneously. Thus, SIMD operations must be used. AVX (Advanced Vector eXtensions) of Intel support SIMD. As the memory operation instructions of AVX used in L4 Pointer are carried out atomically, the L4 Pointer supports multithread safety in Intel architectures. However, in the ARM architecture, certain conditions must be satisfied to ensure that the SIMD instruction is atomic. ARM only provides SIMD atomic when the SIMD element is 32-bit or small and aligned. Since the prototype of the proposed L4 pointer is Intel architecture, it is not implemented to apply to ARM, but ARM can provide it as well.

\begin{figure}[ht]
  \centering
  \includegraphics[width=\linewidth]{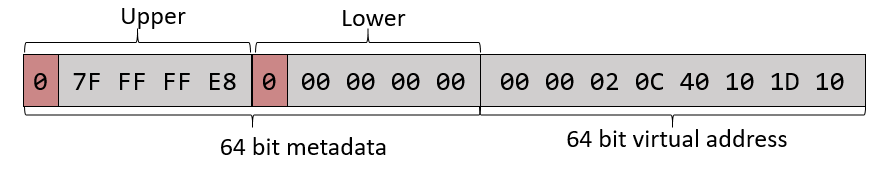}
  \caption{The layout of L4 Pointer.}
  \Description{The layout of L4 Pointer..}
  \label{layout}
\end{figure}

\begin{figure}[ht]
  \centering
  \includegraphics[width=\linewidth]{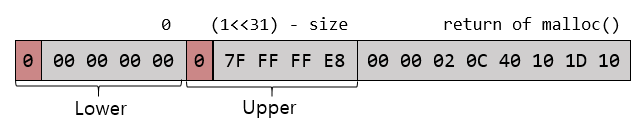}
  \caption{The initialization of the L4 Pointer as "p=malloc(size);".}
  \Description{The layout of L4 Pointer..}
  \label{init}
\end{figure}

\subsection{Layout of L4 Pointer and initialization}
The layout of the L4 Pointer is illustrated in Fig.~\ref{layout}. As previously mentioned, the upper 64 bits of the L4 pointer are used to store metadata, whereas the lower 64 bits are used for the virtual address. The upper 64 bits are halved, the upper 32 bits are used for upper-bound storage, and the lower 32 bits are used for lower-bound storage. In each of the 32 bits, bound information is stored in only 31 bits, and the highest 1 bit is used as a flag. 
 
\begin{equation*}\
stmt\rightarrow expr=expr 
\end{equation*}
\begin{equation*}
\begin{aligned}
expr &\rightarrow expr \ op \ expr \ |\  IDENT_{[constant]}\\
&\rightarrow expr[m:n]
\end{aligned}
\end{equation*}
\begin{equation*}
op \rightarrow +|-|\smallcircle|<<|>>|AND|OR
\end{equation*}

The (1)\verb|~|(12) are expressions of the L4 Pointer expressed as bit vector \cite{Bit-vector}. This paper used the bit-vector formula put forward by C.W.Barrett et al \cite{Bit-vector}. 
The syntax of the bit-vector arithmetic is given below. The subscript [constant] in the expression is the size of the bit vector. 
For example, the subscript "[64]" means that the size of bit vector A is 64. The “[m:n]” means the slicing of the bit vector from m$^{th}$ to n$^{th}$, and m and n are constant. The op and mean multiplication and concatenation of bit vector, respectively.

\begin{equation}
    L4_{[constant]}=pointer_{[64]}\smallcircle upper_{[32]}\smallcircle lower_{[32]}
\end{equation}
\begin{equation}
    pointer_{[64]}=L4_{[128]} [0:63]
\end{equation}
\begin{equation}
    upper_{32}=L4_{[128]}[64:95]
\end{equation}
\begin{equation}
    lower_{[32]}=L4_{[128]}[96:127]
\end{equation}

In the formula in (1), "L4" is a 128-bit L4 pointer and consists of the concatenation of the upper and lower bounds and the real pointer, as the upper, lower, and pointer, respectively. Conversely, the upper and lower are expressed using L4 as (2), (3), and (4), respectively. indicating that it is a slice of L4
The expression, when a pointer is allocated, (5) ~(7) is formulated as below. The pointer is a recorded return of malloc() in (5). The upper is initialized as in the below formula (6), and the reason for this definition is to set the most significant bit to one when a value exceeding size is added. The lower bit is set to zero. Similarly, when the negative offset is added to the initial value, the most significant bit is set to one. Fig.~\ref{init} is the initialization of the process (5)~(7)

\begin{equation}
    pointer_{[64]}=returnOfMalloc_{[64]}
\end{equation}
\begin{equation}
    upper_{[32]}=(1<<32)_{[32]}-size_{[32]}
\end{equation}
\begin{equation}
    lower_{[32]}=0
\end{equation}

\begin{figure}[ht]
  \centering
  \includegraphics[width=\linewidth]{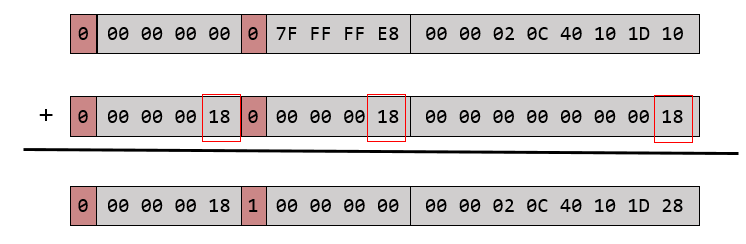}
  \caption{The arithmetic of L4 Pointer as "p+=0x18;".}
  \Description{The layout of L4 Pointer..}
\end{figure} 
\label{arithmetic}
\subsection{Pointer Arithmetic}
The below expressions are the process of pointer arithmetic. 
When the pointer is actually calculated, the offset is added from the base pointer as shown in the formula (8) below.  
The offset is not added only to but is also added to the  and   respectively. 
Fig.~\ref{arithmetic} is the arithmetic of L4 Pointer as the explanation.
Although it looks like each is added and combined in an expression, it is actually made up of one instruction using SIMD operation. 
The details are below section. In summary, the formula is as follows.  

\begin{equation}
\begin{aligned}
    pointer+offset&=L4_{[128]}+offset_{[64]}\smallcircle offset_{[32]}\\
    &\smallcircle offset_{[32]}
\end{aligned}
\end{equation}
First, the following expressions (10) and (11) are performed so 
that only the most significant bit of the lower and upper is left and other bits are set to zero. Next, if the most significant bit of the lower or upper is one, a bit vector of which the most significant bit is set to one, called signal in (14), is created. Then, OR is performed with a pointer as shown in (15). The result will be used as the operand of the load or store. 

\subsection{Bounds check}
The method used in this paper utilizes the exception of memory management units (MMU) without inserting a condition jump for bounds check. Most existing studies have a condition jump as “if statements” for bounds check. Bounds check without a condition jump is used for the delta pointer \cite{Delta}. Pointers must be in the canonical form on Intel architectures. If it is not in canonical form, the MMU (Memory Management Units) of Intel architectures raises an exception. It is possible to terminate the program using the exception that occurred at this time. Setting the most significant bit to one breaks the canonical form when a buffer overflow or underflow occurs. This method can also be used in ARM architecture, where the most significant bits [63:47] must be all zeros or ones. Other than that, an exception is raised for the bit. Therefore, if the most significant bit of pointers is set to one, the MMU of Intel and ARM architectures raise an exception. 

\begin{figure}[t]
  \centering
  \includegraphics[width=\linewidth]{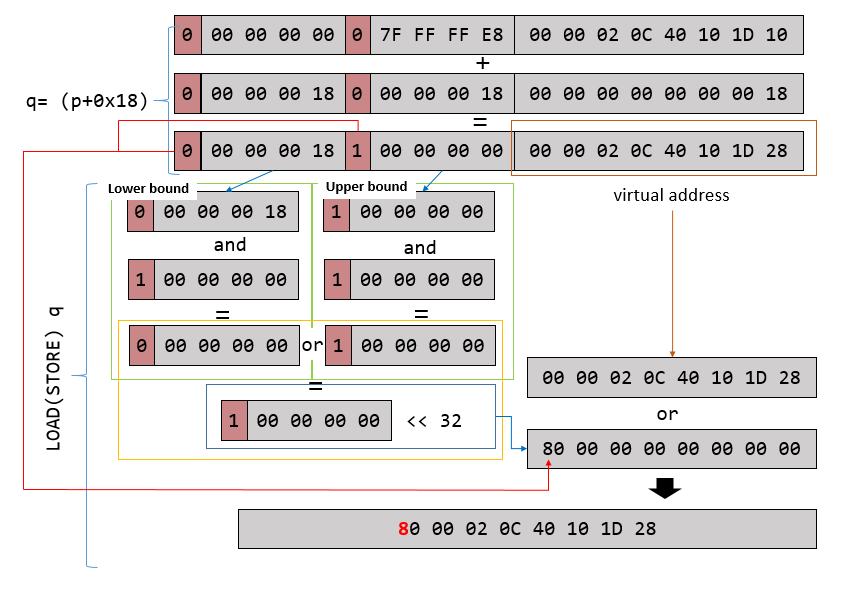}
  \caption{The example of operation of L4 Pointer. The top is pointer arithmetic, the bottom (left) is the tag clear for dereferencing pointers, bottom (right) is making a virtual address for dereferencing pointers. }
  \Description{The layout of L4 Pointer..}
  \label{exmaple}
\end{figure}

If the pointer is exceeded, the most significant bit is set to one. Most architectures do not utilize as many memory addresses as 2\textsuperscript{64} (there are differences between architectures). Therefore, if the most significant bit is set to one, an exception is raised in the CPU’s MMU and the program is terminated. 

\begin{equation}
    msb\textunderscore lower_{[32]} = lower_{[32]} \ AND \ (1<<31)_{[32]}
\end{equation}
\begin{equation}
    msb\textunderscore upper_{[32]} = upper_{[32]}\ AND \ (1<<31)_{[32]}
\end{equation}
\begin{equation}
\begin{aligned}
    msb\textunderscore bits_{[64]} = &(0_{[32]} \ \smallcircle  (msb\textunderscore lower_{[32]} \ OR \ msb\textunderscore upper_{[32]})) \\ & OR \ (1<< 63)_{[64]} 
\end{aligned}
\end{equation}
\begin{equation}
    msb\textunderscore bits_{[64]} = pointer_{[64]} \ OR \ msb\textunderscore bits_{[64]}
\end{equation}

\subsection{Example of manipulations of L4 Pointer}
Fig. \ref{exmaple} is an example of a pointer arithmetic and dereferencing pointer. For the L4 Pointer arithmetic, the offset converts to the L4 Pointer. The locations of the pointer, upper and lower bound are set to offset. To leave only the most significant bit of the upper and lower bounds, we perform an ’AND’ operation on 0x80000000, upper and lower bounds, and upper and lower bounds combined. We perform an ’OR’ operation on the result and virtual address. The result of the operation is dereferencing the pointer. The buffer overflow occurs in Fig. \ref{exmaple}, so the most significant bit is set to one, so MMU (Memory Management Unit) raises an exception and terminates the program.

\section{Transformations and Implementations}
\subsection{Type conversions}
The L4 pointer was implemented as code instrumentation using LLVM pass \cite{LLVM}. All pointer-type variables in the program convert to the L4 pointer. As per the code below, the pointer type is converted to L4 type. The L4 is wrapping a pointer type. Unwrapping the L4 will return the pointer type. 

\begin{equation}
\label{type_conversions}
int* ptr; \rightarrow L4_{<int*>} ptr;
\end{equation}

The pointer array is also converted to an array of L4 type array. It will change as shown below. The size of the pointer type is 64 bits, so the size of the array increases because L4 is 128 bits.  \\
\begin{equation}
int* ptr[5]; \rightarrow L4_{<int*>} ptr[5];
\end{equation}

The type of the member of the structure is a pointer, it is converted to L4. Therefore, the argument of the allocation function such as malloc must be changed. The argument is changed according to the size of the changed structure. 


\subsection{The way to protect stack and global object}
L4 The pointer can checks the bound of the stack and global objects as well. In C programs, they are an array. In order to support the array, it is converted to indirect access to an array. If an array is declared, an L4 type variable is additionally declared, and the address and metadata information of the array are stored in the L4 type. Then, if there is access to the array, it
\begin{equation}
int ptr[10];  
\end{equation}
\begin{equation}
L4_{<int*>} indirectPtr = makeL4(\verb|&|ptr, 10);    
\end{equation}


\begin{table}[]
\caption{The example code of L4 Pointer like C}
\resizebox{\columnwidth}{!}{%
\begin{tabular}{ll}
1.  & \cellcolor[HTML]{C0C0C0}struct L4\{                                                                         \\
2.  & \cellcolor[HTML]{C0C0C0}  uint64\_t tag;                                                                     \\
3.  & \cellcolor[HTML]{C0C0C0}  uint64\_t ptr;                                                                     \\
4.  & \cellcolor[HTML]{C0C0C0}\};                                                                                \\
5.  & void foo(char* ptr)                                                                                        \\
6.  & \cellcolor[HTML]{C0C0C0}void foo(L4 ptr)\{                                                                 \\
7.  &     ptr{[}3{]} = ‘a’;                                                                                          \\
8.  & \cellcolor[HTML]{C0C0C0}  uint128\_t tmp= (uint128\_t) ptr;                                                  \\
9.  & \cellcolor[HTML]{C0C0C0}  tmp+= (index \textless{}\textless 96 )| (index\textless{}\textless{}64) | (index); \\
10  & \cellcolor[HTML]{C0C0C0}  L4 temp\_ptr= (L4) tmp;                                                            \\
11. & \cellcolor[HTML]{C0C0C0}  uint64\_t upper\_tag= temp\_ptr\& 0x8000000000;                                    \\
12. & \cellcolor[HTML]{C0C0C0}  uint64\_t lower\_tag= (temp\_ptr\textless{}\textless 32) \& 0x8000000000;          \\
13. & \cellcolor[HTML]{C0C0C0}  uint64\_t tag = upper\_tag| lower\_tag;                                            \\
14. & \cellcolor[HTML]{C0C0C0}  char* accessed\_ptr= temp\_ptr.ptr| tag;                                           \\
15. & \cellcolor[HTML]{C0C0C0}  accessed\_ptr{[}3{]} = 'a';                                                        \\
16. & \}                                                                                                         \\
17. &int main()\{                                                                                               \\
18. &   char* ptr;                                                                                                 \\
19  & \cellcolor[HTML]{C0C0C0}  L4 L4\_ptr;                                                                        \\
20. & ptr= malloc(100);                                                                                          \\
21. & \cellcolor[HTML]{C0C0C0}  address = malloc(100);                                                             \\
22. & \cellcolor[HTML]{C0C0C0}  L4\_tag = ((1 \textless{}\textless 31 ) – 100) \textless{}\textless 32;            \\
23. & \cellcolor[HTML]{C0C0C0}  L4\_ptr.tag = L4\_tag;                                                             \\
24. & \cellcolor[HTML]{C0C0C0}  L4\_ptr.ptr = address;                                                             \\
25. & foo(ptr);                                                                                                  \\
26.  & \cellcolor[HTML]{C0C0C0}  foo(L4\_ptr);                                                                      \\
27. & \}                                                                                                        
\end{tabular}%
}
\label{examplecode}
\end{table}

\subsection{Examples}
Table~\ref{examplecode} is an example of when the source code of the target program is instrumented by the L4 Pointer. The white codes are the original codes, and the gray code is the instrumented code. A structure called L4 is added for readability, but it is done by vector operations. When the ‘malloc()’ is called, the code in lines 21 to 24 is performed to make an L4 pointer. If the argument of the function is a pointer, the transformation of the function is also performed to pass the L4 Pointer. 
The pointer arithmetic operations use special registers. The prototype of the L4 Pointer is currently operating on an Intel x86 \cite{Intel}. Special registers called the XMM registers of Intel x86 are 128-bit long and used for vector or float calculations. Fig.~\ref{asm} shows the pointer operation process in the x86 instruction. The instructions ‘movaps’ and ‘paddq’ are used instead of ‘mov’ and ‘add’ and are used for the XMM registers. The calculated value was then set in the XMM0 register, and the next operation is performed using the ‘paddq’ instruction. The result of the operation is stored in the XMM1 register and memory.

\begin{figure}[ht]
\begin{verbatim}
movaps XMM1, XMMWORD PTR [rbp-0x20]
movaps XMM0, 0x1000000010000000000000001   
paddq  XMM1, XMM0
movaps XMMWORD PTR [rbp-0x30], XMM1
\end{verbatim}
\caption{Example assembly codes in Intel x86 Architecture used L4 Pointer }
\label{asm}
\end{figure}

\subsection{Implementations}
The L4 pointer is implemented in the form of Pass of LLVM \cite{LLVM}. We used WLLVM to transform the code \cite{wllvm}. WLLVM merges multiple source codes into one bitcode, and enables the transformation of the code into a single file. The reason for making one-bit code is to ensure correspondence to the structure. As mentioned earlier, pointers in structures are also converted to L4 pointers. External library calls are made by changing these to general pointers. In order to reduce external source code calls to maintain the L4 pointer, they are combined into one file and transformed. The 128-bit pointer type is implemented as a vector type. The target of the prototype is x86, and it is implemented as a type of <2 * i64>. As mentioned earlier, the pointer is converted to a vector type and if ARM is targeted, the type of the L4 pointer becomes <4 * i32> so that atomicity can be guaranteed. 


\section{Evaluation}
This paper shows evaluations of the L4 Pointer in terms of performance, given as follows: 
\begin{itemize}
    \item Runtime overhead
    \item Memory overhead
    \item Conflict between L4 operations and programs using SIMD instructions
\end{itemize}
To evaluate runtime overhead and memory overhead, we ran benchmark programs, namely Olden, coremark and two programs of mibench. To evaluate the conflict, we ran programs using SIMD in llvm-test-suite. 

\begin{figure}[ht]
  \centering
  \includegraphics[width=\linewidth]{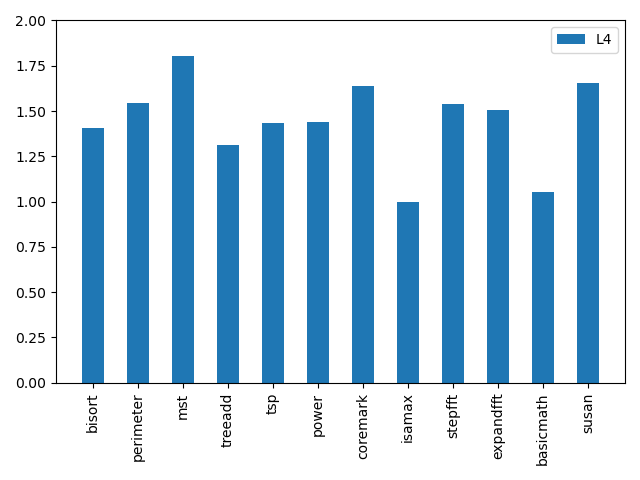}
  \caption{The runtime overhead of L4 Pointer}
  \Description{}
  \label{runtime}
\end{figure}
\begin{figure}[ht]
  \centering
  \includegraphics[width=\linewidth]{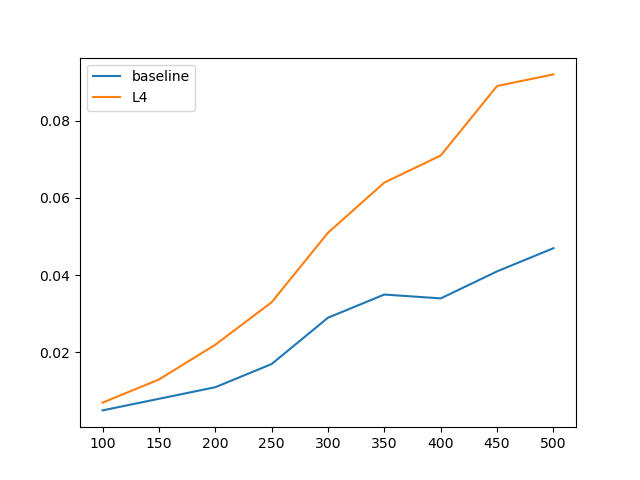}
  \caption{The runtime overhead of mst based on program's input, input is the number of a vertex of graph, the vertex is a linked list structure}
  \Description{}
  \label{mst}
\end{figure}
\begin{figure}[ht]
  \centering
  \includegraphics[width=\linewidth]{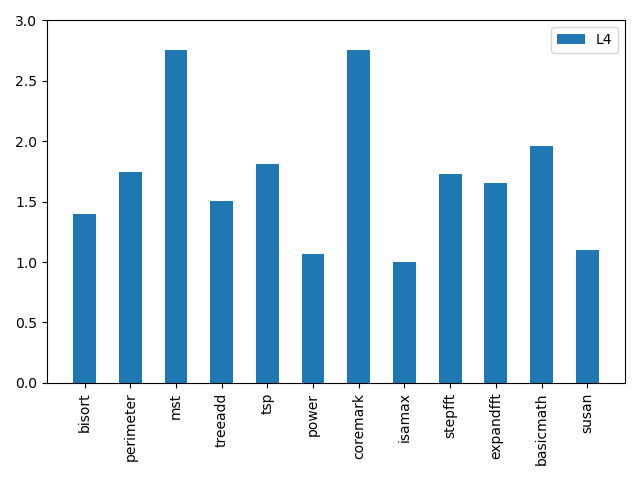}
  \caption{The memory overhead of L4 Pointer}
  \Description{}
  \label{memory}
\end{figure}
\begin{figure}[ht]
  \centering
  \includegraphics[width=\linewidth]{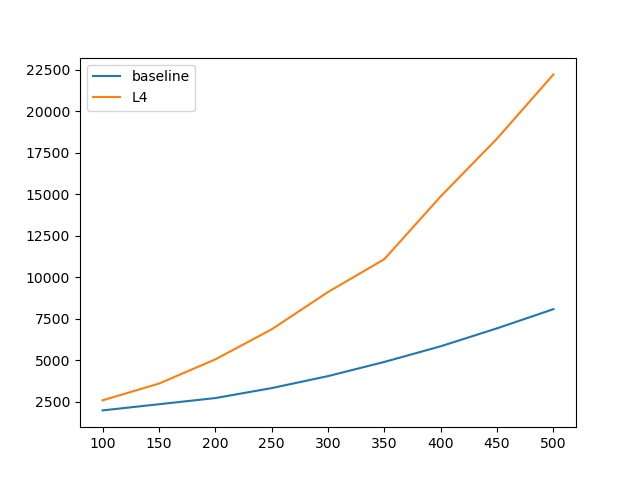}
  \caption{The memory overhead of mst based on program's input}
  \Description{}
  \label{mst_memory}
\end{figure}

\begin{figure}[ht]
  \centering
  \includegraphics[width=\linewidth]{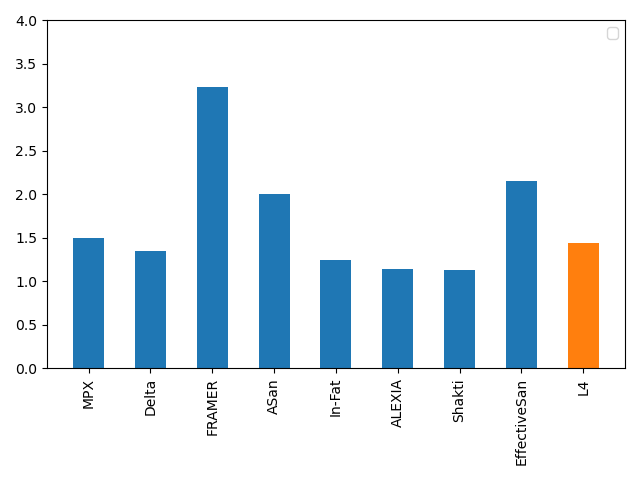}
  \caption{Comparisons between related works and L4 Pointer}
  \Description{}
  \label{comprarisons}
\end{figure}

\subsection{Runtime overhead}
To test the runtime overhead, we used three benchmarks, namely olden, coremark, and mibench, and test cases of llvm-test suite \cite{Olden, coremark, Mibench}. Six programs of olden are executed, one of coremark, and two of mibench, viz. basicmath and susan. In the llvm-test-suite, isamax, stepfft, and expandfft were executed. As will be explained in the next subsection, this test case was tested to check whether there was a conflict with the SIMD program. 

Fig.~\ref{runtime} is the graph of runtime overhead of the L4 Pointer, and the slow-down of the L4 Pointer averaged 2.07×. The average overhead is lower than the reported values from FRAMER (3.23×). It is slower than the delta pointer, which is 1.3×, but the delta pointer only checks the upper bound. The bounds
heck of EffectiveSan incurs overhead about 2.15× as slow as ours, but only supports sub-objects. L4 pointer is slower than the hardware-assisted approach. In-Fat, ALEXIA, and Shakti-MS incur 1.24×, 1.14×, and 1.13×, respectively. However, L4 pointer has been implemented without any architectural extensions. 
The highest overhead is 1.8x for mst in olden benchmark. 
Mst uses a kind of a linked list, which uses multiple nonspatial pointers in a vertex, but all are converted to L4 Pointers, causing a slow down.

Fig.~\ref{runtime} depicts the runtime overhead of L4 Pointer, and the slowdown of L4 Pointer averaged 1.44x. The average overhead is lower than the reported values from FRAMER (3.23x). Delta pointer is 1.3x, it is slower than delta pointer, but delta pointer only checks the upper bound. 
The bounds check of Effective-San incurs overhead about 2.15x as slow as ours, but only supports subobjects.
L4 pointer is slower than the hardware-assisted approach. Intel MPX, In-Fat, ALEXIA, and Shakti-MS incur 1.5x, 1.24x, 1.14x, and 1.13x, respectively. However, L4 pointer has been implemented without any architectural extensions.
The highest overhead is 1.8x for mst in olden benchmark, again. 

Experiments were conducted with various inputs of mst. 
In the graph of Fig.~\ref{mst}, depending on the input of mst, the number of vertices is the size of the linked list structure. 
It was confirmed that the larger the input, the closer to twice.
As will be explained later, the runtime overhead increases up to twice, but the memory overhead increases up to three times. 
With the bounds checking code added for each vertex, the instruction is executed linearly along the linked list, yielding runtime overhead in proportion to the number of vertices.
On the other hand, memory overhead increases faster than runtime overhead along the size of the linked list. 
We guess that it is because each vertex has more than one pointer field and 
the extra memory for all the vertices  
collectively forms a kind of tree, instead of a linked list.


\
\subsection{Memory overhead}
We measure memory overhead using the same programs mentioned in the above subsection. Fig.~\ref{memory} is the graph of the memory overhead of the L4 pointer. The memory overhead averaged 1.7×. Address Sanitizer and Memory Sanitizer incur memory overheads of 3.37× and 3.32× in their paper. The memory overhead of FRAMER averaged 1.23× smaller than ours because L4 Pointer is twice the size of the normal pointer. The In-Fat pointer incurs memory overhead of 1.21× in their paper. Mst and coremark are higher memory overhead than others, they use a linked list. As mentioned above, a linked list also incurs memory overhead. 
As with runtime overhead, mst recorded high memory overhead. Similarly, when the input was changed, the memory overhead increased as the number of vertices increased. Fig.~\ref{mst_memory} is the result. Unlike runtime overhead convergence to 2x, memory overhead increased from 2x to 3x. It was confirmed that the memory overhead increases as the number of the used linked lists increases.
"Vertex" structure type in mst has two pointers as members. Since an L4 pointer is twice the size of a pointer, a structure with many pointer members takes a lot of memory overhead. In addition, there are many pointer types as members in other used structure types.

\subsection{The programs using SIMD}
In Fig.~\ref{runtime} and Fig.~\ref{memory}, isamax, stepfft and expandfft are programs using SIMD and which are instrumented and executed by L4 Pointer. They worked without conflict and the average of runtime and memory overhead is 1.34× and 1.46×, respectively. AVX of Intel supports XMM0–XMM15 registers and L4 Pointer only uses "movaps" and "paddq" \cite{Intel}. Therefore, L4 Pointer does not conflict with SIMD programs.

\section{Disscusion}
As previously mentioned, Table~\ref{comparisonsofrelatedworks} shows a comparison between related works and the L4 Pointer, and Fig.~\ref{comprarisons} is a graph that compares performance. On the graph, L4 Pointer has a lower runtime overhead than FRAMER. FRAMER supports full bounds checking but has the limitation of being software-only. Delta pointer requires a lower runtime overhead than L4 Pointer, but delta pointer only supports upper bound. EffectiveSan only subobject bounds check. MPX does not support multithread safety. ASan incurs twice of runtime overhead mentioned in their paper, but four times of runtime overhead. Hardware-assisted approaches have a lower runtime overhead than L4 Pointer, however, they are not available on the existing architectures. 
That is, they need extended hardware, to support multithread safety as well as lower overhead bounds checking using instruction set extensions, and multithread safety. 
We could not compare L4 Pointer with other studies for memory overhead 
because most of them did not publicize memory overhead.
Only the In-Fat pointer incurs 1.21 times memory overhead, which is lower than L4 Pointer. 

Experiments show that 
the more overhead occurs,
the more pointer fields a structural object has.
Therefore, many studies do not insert bounds checking for pointers embedded in a structure object, to minimize the number of bounds checking, and improve performance \cite{softbound, ABCD, ArrayBondsCheck}. 
However,  we insert bounds checking into all kinds of pointers, in order not to lose complicated cases of buffer overflow.
In the future, we will study how to improve performance while ensuring safety for structures with multiple pointer fields.

\section{Conclusions}
This paper proposes new bounds checking named L4 Pointer. L4 pointer is applicable to any architecture that provides SIMD operation. By using SIMD Operation, multithread safety is ensured, and sufficient metadata storage space is provided. L4 pointer was 1.44× slow-down without hardware extensions, and memory overhead was 1.7×.

\section{Acknowledgments}
We would like to thank Editage (www.editage.co.kr) for English language editing. This work was supported by Institute for Information \verb|&| communications Technology Planning \verb|&| Evaluation(IITP) grant funded by the Korea government (MSIT)(No.2022-0-01200, Training Key Talents in Industrial Convergence Security)
\bibliographystyle{ACM-Reference-Format}
\bibliography{main}


\begin{thebibliography}{33}


\ifx \showCODEN    \undefined \def \showCODEN     #1{\unskip}     \fi
\ifx \showDOI      \undefined \def \showDOI       #1{#1}\fi
\ifx \showISBNx    \undefined \def \showISBNx     #1{\unskip}     \fi
\ifx \showISBNxiii \undefined \def \showISBNxiii  #1{\unskip}     \fi
\ifx \showISSN     \undefined \def \showISSN      #1{\unskip}     \fi
\ifx \showLCCN     \undefined \def \showLCCN      #1{\unskip}     \fi
\ifx \shownote     \undefined \def \shownote      #1{#1}          \fi
\ifx \showarticletitle \undefined \def \showarticletitle #1{#1}   \fi
\ifx \showURL      \undefined \def \showURL       {\relax}        \fi
\providecommand\bibfield[2]{#2}
\providecommand\bibinfo[2]{#2}
\providecommand\natexlab[1]{#1}
\providecommand\showeprint[2][]{arXiv:#2}

\bibitem[Int(2016)]%
        {Intel}
Intel Corporation \bibinfo{year}{2016}\natexlab{}.
\newblock \bibinfo{booktitle}{\emph{Intel® 64 and IA-32 Architectures Software
  Developer’s Manual}}.
\newblock Intel Corporation.
\newblock
\urldef\tempurl%
\url{https://www.intel.com/content/dam/www/public/us/en/documents/manuals/64-ia-32-architectures-software-developer-vol-1-manual.pdf}
\showURL{%
\tempurl}


\bibitem[ARM(2020)]%
        {ARM}
Arm Limited \bibinfo{year}{2020}\natexlab{}.
\newblock \bibinfo{booktitle}{\emph{1. ARM Architecture Reference Manual
  ARMv8}}.
\newblock Arm Limited.
\newblock
\urldef\tempurl%
\url{https://developer.arm.com/documentation/ddi0487/latest}
\showURL{%
\tempurl}


\bibitem[cor(line)]%
        {coremark}
EEMBC Certification Lab \bibinfo{year}{2022 [online]}\natexlab{}.
\newblock \bibinfo{booktitle}{\emph{coremark}}.
\newblock EEMBC Certification Lab.
\newblock
\urldef\tempurl%
\url{https://www.eembc.org/coremark/}
\showURL{%
Retrieved November 11, 2022 from \tempurl}


\bibitem[Akritidis et~al\mbox{.}(2009)]%
        {Baggybounds}
\bibfield{author}{\bibinfo{person}{Periklis Akritidis}, \bibinfo{person}{Manuel
  Costa}, \bibinfo{person}{Miguel Castro}, {and} \bibinfo{person}{Steven
  Hand}.} \bibinfo{year}{2009}\natexlab{}.
\newblock \showarticletitle{Baggy Bounds Checking: An Efficient and
  Backwards-Compatible Defense against out-of-Bounds Errors}
  \emph{(\bibinfo{series}{SSYM'09})}. \bibinfo{publisher}{USENIX Association},
  \bibinfo{address}{USA}, \bibinfo{pages}{51–66}.
\newblock


\bibitem[Barrett et~al\mbox{.}(1998)]%
        {Bit-vector}
\bibfield{author}{\bibinfo{person}{Clark~W. Barrett}, \bibinfo{person}{David~L.
  Dill}, {and} \bibinfo{person}{Jeremy~R. Levitt}.}
  \bibinfo{year}{1998}\natexlab{}.
\newblock \showarticletitle{A Decision Procedure for Bit-Vector Arithmetic}. In
  \bibinfo{booktitle}{\emph{Proceedings of the 35th Annual Design Automation
  Conference}} (San Francisco, California, USA) \emph{(\bibinfo{series}{DAC
  '98})}. \bibinfo{publisher}{Association for Computing Machinery},
  \bibinfo{address}{New York, NY, USA}, \bibinfo{pages}{522–527}.
\newblock
\showISBNx{0897919645}
\urldef\tempurl%
\url{https://doi.org/10.1145/277044.277186}
\showDOI{\tempurl}


\bibitem[Bod\'{\i}k et~al\mbox{.}(2000)]%
        {ABCD}
\bibfield{author}{\bibinfo{person}{Rastislav Bod\'{\i}k},
  \bibinfo{person}{Rajiv Gupta}, {and} \bibinfo{person}{Vivek Sarkar}.}
  \bibinfo{year}{2000}\natexlab{}.
\newblock \showarticletitle{ABCD: Eliminating Array Bounds Checks on Demand}.
\newblock \bibinfo{journal}{\emph{SIGPLAN Not.}} \bibinfo{volume}{35},
  \bibinfo{number}{5} (\bibinfo{date}{may} \bibinfo{year}{2000}),
  \bibinfo{pages}{321–333}.
\newblock
\showISSN{0362-1340}
\urldef\tempurl%
\url{https://doi.org/10.1145/358438.349342}
\showDOI{\tempurl}


\bibitem[Carlisle and Rogers(1995)]%
        {Olden}
\bibfield{author}{\bibinfo{person}{Martin~C. Carlisle} {and}
  \bibinfo{person}{Anne Rogers}.} \bibinfo{year}{1995}\natexlab{}.
\newblock \showarticletitle{Software Caching and Computation Migration in
  Olden}.
\newblock \bibinfo{journal}{\emph{SIGPLAN Not.}} \bibinfo{volume}{30},
  \bibinfo{number}{8} (\bibinfo{date}{aug} \bibinfo{year}{1995}),
  \bibinfo{pages}{29–38}.
\newblock
\showISSN{0362-1340}
\urldef\tempurl%
\url{https://doi.org/10.1145/209937.209941}
\showDOI{\tempurl}


\bibitem[Das et~al\mbox{.}(2019)]%
        {SHAKTI}
\bibfield{author}{\bibinfo{person}{Sourav Das},
  \bibinfo{person}{R.~Harikrishnan Unnithan}, \bibinfo{person}{Arjun Menon},
  \bibinfo{person}{Chester Rebeiro}, {and} \bibinfo{person}{Kamakoti
  Veezhinathan}.} \bibinfo{year}{2019}\natexlab{}.
\newblock \showarticletitle{SHAKTI-MS: A RISC-V Processor for Memory Safety in
  C} \emph{(\bibinfo{series}{LCTES 2019})}. \bibinfo{publisher}{Association for
  Computing Machinery}, \bibinfo{address}{New York, NY, USA},
  \bibinfo{pages}{19–32}.
\newblock
\showISBNx{9781450367240}
\urldef\tempurl%
\url{https://doi.org/10.1145/3316482.3326356}
\showDOI{\tempurl}


\bibitem[DATABASE(2020)]%
        {underflow2}
\bibfield{author}{\bibinfo{person}{NATIONAL~VULNERABILITY DATABASE}.}
  \bibinfo{year}{2020}\natexlab{}.
\newblock \bibinfo{booktitle}{\emph{CVE-2020-24658 Detail}}.
\newblock
\urldef\tempurl%
\url{https://nvd.nist.gov/vuln/detail/CVE-2020-24658}
\showURL{%
Retrieved November 11, 2022 from \tempurl}


\bibitem[Devietti et~al\mbox{.}(2008)]%
        {Hardbound}
\bibfield{author}{\bibinfo{person}{Joe Devietti}, \bibinfo{person}{Colin
  Blundell}, \bibinfo{person}{Milo M.~K. Martin}, {and} \bibinfo{person}{Steve
  Zdancewic}.} \bibinfo{year}{2008}\natexlab{}.
\newblock \showarticletitle{Hardbound: Architectural Support for Spatial Safety
  of the C Programming Language} \emph{(\bibinfo{series}{ASPLOS XIII})}.
  \bibinfo{publisher}{Association for Computing Machinery},
  \bibinfo{address}{New York, NY, USA}, \bibinfo{pages}{103–114}.
\newblock
\showISBNx{9781595939586}
\urldef\tempurl%
\url{https://doi.org/10.1145/1346281.1346295}
\showDOI{\tempurl}


\bibitem[Dhurjati and Adve(2006)]%
        {Backwards}
\bibfield{author}{\bibinfo{person}{Dinakar Dhurjati} {and}
  \bibinfo{person}{Vikram Adve}.} \bibinfo{year}{2006}\natexlab{}.
\newblock \showarticletitle{Backwards-Compatible Array Bounds Checking for C
  with Very Low Overhead} \emph{(\bibinfo{series}{ICSE '06})}.
  \bibinfo{publisher}{Association for Computing Machinery},
  \bibinfo{address}{New York, NY, USA}, \bibinfo{pages}{162–171}.
\newblock
\showISBNx{1595933751}
\urldef\tempurl%
\url{https://doi.org/10.1145/1134285.1134309}
\showDOI{\tempurl}


\bibitem[Dhurjati et~al\mbox{.}(2003)]%
        {memorysafety}
\bibfield{author}{\bibinfo{person}{Dinakar Dhurjati}, \bibinfo{person}{Sumant
  Kowshik}, \bibinfo{person}{Vikram Adve}, {and} \bibinfo{person}{Chris
  Lattner}.} \bibinfo{year}{2003}\natexlab{}.
\newblock \showarticletitle{Memory Safety without Runtime Checks or Garbage
  Collection}. In \bibinfo{booktitle}{\emph{Proceedings of the 2003 ACM SIGPLAN
  Conference on Language, Compiler, and Tool for Embedded Systems}} (San Diego,
  California, USA) \emph{(\bibinfo{series}{LCTES '03})}.
  \bibinfo{publisher}{Association for Computing Machinery},
  \bibinfo{address}{New York, NY, USA}, \bibinfo{pages}{69–80}.
\newblock
\showISBNx{1581136471}
\urldef\tempurl%
\url{https://doi.org/10.1145/780732.780743}
\showDOI{\tempurl}


\bibitem[Duck and Yap(2018)]%
        {EffectiveSan}
\bibfield{author}{\bibinfo{person}{Gregory~J. Duck} {and}
  \bibinfo{person}{Roland H.~C. Yap}.} \bibinfo{year}{2018}\natexlab{}.
\newblock \showarticletitle{EffectiveSan: Type and Memory Error Detection Using
  Dynamically Typed C/C++}.
\newblock \bibinfo{journal}{\emph{SIGPLAN Not.}} \bibinfo{volume}{53},
  \bibinfo{number}{4} (\bibinfo{date}{jun} \bibinfo{year}{2018}),
  \bibinfo{pages}{181–195}.
\newblock
\showISSN{0362-1340}
\urldef\tempurl%
\url{https://doi.org/10.1145/3296979.3192388}
\showDOI{\tempurl}


\bibitem[Durumeric et~al\mbox{.}(2014)]%
        {heartbleed}
\bibfield{author}{\bibinfo{person}{Zakir Durumeric}, \bibinfo{person}{Frank
  Li}, \bibinfo{person}{James Kasten}, \bibinfo{person}{Johanna Amann},
  \bibinfo{person}{Jethro Beekman}, \bibinfo{person}{Mathias Payer},
  \bibinfo{person}{Nicolas Weaver}, \bibinfo{person}{David Adrian},
  \bibinfo{person}{Vern Paxson}, \bibinfo{person}{Michael Bailey}, {and}
  \bibinfo{person}{J.~Alex Halderman}.} \bibinfo{year}{2014}\natexlab{}.
\newblock \showarticletitle{The Matter of Heartbleed}. In
  \bibinfo{booktitle}{\emph{Proceedings of the 2014 Conference on Internet
  Measurement Conference}} (Vancouver, BC, Canada) \emph{(\bibinfo{series}{IMC
  '14})}. \bibinfo{publisher}{Association for Computing Machinery},
  \bibinfo{address}{New York, NY, USA}, \bibinfo{pages}{475–488}.
\newblock
\showISBNx{9781450332132}
\urldef\tempurl%
\url{https://doi.org/10.1145/2663716.2663755}
\showDOI{\tempurl}


\bibitem[Guthaus et~al\mbox{.}(2001)]%
        {Mibench}
\bibfield{author}{\bibinfo{person}{M.R. Guthaus}, \bibinfo{person}{J.S.
  Ringenberg}, \bibinfo{person}{D. Ernst}, \bibinfo{person}{T.M. Austin},
  \bibinfo{person}{T. Mudge}, {and} \bibinfo{person}{R.B. Brown}.}
  \bibinfo{year}{2001}\natexlab{}.
\newblock \showarticletitle{MiBench: A free, commercially representative
  embedded benchmark suite}. In \bibinfo{booktitle}{\emph{Proceedings of the
  Fourth Annual IEEE International Workshop on Workload Characterization. WWC-4
  (Cat. No.01EX538)}}. \bibinfo{pages}{3--14}.
\newblock
\urldef\tempurl%
\url{https://doi.org/10.1109/WWC.2001.990739}
\showDOI{\tempurl}


\bibitem[Krishnakumar et~al\mbox{.}(2019)]%
        {ALEXIA}
\bibfield{author}{\bibinfo{person}{Gnanambikai Krishnakumar},
  \bibinfo{person}{Kommuru~Alekhya REDDY}, {and} \bibinfo{person}{Chester
  Rebeiro}.} \bibinfo{year}{2019}\natexlab{}.
\newblock \showarticletitle{ALEXIA: A Processor with Lightweight Extensions for
  Memory Safety}.
\newblock \bibinfo{journal}{\emph{ACM Trans. Embed. Comput. Syst.}}
  \bibinfo{volume}{18}, \bibinfo{number}{6}, Article \bibinfo{articleno}{122}
  (\bibinfo{date}{nov} \bibinfo{year}{2019}), \bibinfo{numpages}{27}~pages.
\newblock
\showISSN{1539-9087}
\urldef\tempurl%
\url{https://doi.org/10.1145/3362064}
\showDOI{\tempurl}


\bibitem[Kroes et~al\mbox{.}(2018)]%
        {Delta}
\bibfield{author}{\bibinfo{person}{Taddeus Kroes}, \bibinfo{person}{Koen
  Koning}, \bibinfo{person}{Erik van~der Kouwe}, \bibinfo{person}{Herbert Bos},
  {and} \bibinfo{person}{Cristiano Giuffrida}.}
  \bibinfo{year}{2018}\natexlab{}.
\newblock \showarticletitle{Delta Pointers: Buffer Overflow Checks without the
  Checks}. In \bibinfo{booktitle}{\emph{Proceedings of the Thirteenth EuroSys
  Conference}} (Porto, Portugal) \emph{(\bibinfo{series}{EuroSys '18})}.
  \bibinfo{publisher}{Association for Computing Machinery},
  \bibinfo{address}{New York, NY, USA}, Article \bibinfo{articleno}{22},
  \bibinfo{numpages}{14}~pages.
\newblock
\showISBNx{9781450355841}
\urldef\tempurl%
\url{https://doi.org/10.1145/3190508.3190553}
\showDOI{\tempurl}


\bibitem[Lattner and Adve(2004)]%
        {LLVM}
\bibfield{author}{\bibinfo{person}{Chris Lattner} {and} \bibinfo{person}{Vikram
  Adve}.} \bibinfo{year}{2004}\natexlab{}.
\newblock \showarticletitle{LLVM: A Compilation Framework for Lifelong Program
  Analysis \& Transformation} \emph{(\bibinfo{series}{CGO '04})}.
  \bibinfo{publisher}{IEEE Computer Society}, \bibinfo{address}{USA},
  \bibinfo{pages}{75}.
\newblock
\showISBNx{0769521029}


\bibitem[MITRE(2022a)]%
        {CWE}
\bibfield{author}{\bibinfo{person}{MITRE}.} \bibinfo{year}{2022}\natexlab{a}.
\newblock \bibinfo{booktitle}{\emph{Common Weakness Enumeration}}.
\newblock
\urldef\tempurl%
\url{https://cwe.mitre.org/}
\showURL{%
Retrieved November 24, 2022 from \tempurl}


\bibitem[MITRE(2022b)]%
        {CWETOP25}
\bibfield{author}{\bibinfo{person}{MITRE}.} \bibinfo{year}{2022}\natexlab{b}.
\newblock \bibinfo{booktitle}{\emph{CWE top 25}}.
\newblock
\urldef\tempurl%
\url{https://cwe.mitre.org/top25/archive/2022/2022_cwe_top25.html}
\showURL{%
Retrieved November 24, 2022 from \tempurl}


\bibitem[Mullen and Meany(2019)]%
        {IoTBuffer}
\bibfield{author}{\bibinfo{person}{Gary Mullen} {and} \bibinfo{person}{Liam
  Meany}.} \bibinfo{year}{2019}\natexlab{}.
\newblock \showarticletitle{Assessment of Buffer Overflow Based Attacks On an
  IoT Operating System}. In \bibinfo{booktitle}{\emph{2019 Global IoT Summit
  (GIoTS)}}. \bibinfo{pages}{1--6}.
\newblock
\urldef\tempurl%
\url{https://doi.org/10.1109/GIOTS.2019.8766434}
\showDOI{\tempurl}


\bibitem[Nagarakatte et~al\mbox{.}(2009)]%
        {softbound}
\bibfield{author}{\bibinfo{person}{Santosh Nagarakatte},
  \bibinfo{person}{Jianzhou Zhao}, \bibinfo{person}{Milo~M.K. Martin}, {and}
  \bibinfo{person}{Steve Zdancewic}.} \bibinfo{year}{2009}\natexlab{}.
\newblock \showarticletitle{SoftBound: Highly Compatible and Complete Spatial
  Memory Safety for c}. In \bibinfo{booktitle}{\emph{Proceedings of the 30th
  ACM SIGPLAN Conference on Programming Language Design and Implementation}}
  (Dublin, Ireland) \emph{(\bibinfo{series}{PLDI '09})}.
  \bibinfo{publisher}{Association for Computing Machinery},
  \bibinfo{address}{New York, NY, USA}, \bibinfo{pages}{245–258}.
\newblock
\showISBNx{9781605583921}
\urldef\tempurl%
\url{https://doi.org/10.1145/1542476.1542504}
\showDOI{\tempurl}


\bibitem[Nam et~al\mbox{.}(2019)]%
        {FRAMER}
\bibfield{author}{\bibinfo{person}{Myoung~Jin Nam}, \bibinfo{person}{Periklis
  Akritidis}, {and} \bibinfo{person}{David~J Greaves}.}
  \bibinfo{year}{2019}\natexlab{}.
\newblock \showarticletitle{FRAMER: A Tagged-Pointer Capability System with
  Memory Safety Applications} \emph{(\bibinfo{series}{ACSAC '19})}.
  \bibinfo{publisher}{Association for Computing Machinery},
  \bibinfo{address}{New York, NY, USA}, \bibinfo{pages}{612–626}.
\newblock
\showISBNx{9781450376280}
\urldef\tempurl%
\url{https://doi.org/10.1145/3359789.3359799}
\showDOI{\tempurl}


\bibitem[Necula et~al\mbox{.}(2005)]%
        {CCured}
\bibfield{author}{\bibinfo{person}{George~C. Necula}, \bibinfo{person}{Jeremy
  Condit}, \bibinfo{person}{Matthew Harren}, \bibinfo{person}{Scott McPeak},
  {and} \bibinfo{person}{Westley Weimer}.} \bibinfo{year}{2005}\natexlab{}.
\newblock \showarticletitle{CCured: Type-Safe Retrofitting of Legacy Software}.
\newblock \bibinfo{journal}{\emph{ACM Trans. Program. Lang. Syst.}}
  \bibinfo{volume}{27}, \bibinfo{number}{3} (\bibinfo{date}{may}
  \bibinfo{year}{2005}), \bibinfo{pages}{477–526}.
\newblock
\showISSN{0164-0925}
\urldef\tempurl%
\url{https://doi.org/10.1145/1065887.1065892}
\showDOI{\tempurl}


\bibitem[Oleksenko et~al\mbox{.}(2018)]%
        {intelMPX}
\bibfield{author}{\bibinfo{person}{Oleksii Oleksenko}, \bibinfo{person}{Dmitrii
  Kuvaiskii}, \bibinfo{person}{Pramod Bhatotia}, \bibinfo{person}{Pascal
  Felber}, {and} \bibinfo{person}{Christof Fetzer}.}
  \bibinfo{year}{2018}\natexlab{}.
\newblock \showarticletitle{Intel MPX Explained: A Cross-Layer Analysis of the
  Intel MPX System Stack}. In \bibinfo{booktitle}{\emph{Abstracts of the 2018
  ACM International Conference on Measurement and Modeling of Computer
  Systems}} (Irvine, CA, USA) \emph{(\bibinfo{series}{SIGMETRICS '18})}.
  \bibinfo{publisher}{Association for Computing Machinery},
  \bibinfo{address}{New York, NY, USA}, \bibinfo{pages}{111–112}.
\newblock
\showISBNx{9781450358460}
\urldef\tempurl%
\url{https://doi.org/10.1145/3219617.3219662}
\showDOI{\tempurl}


\bibitem[OpenSSL~Foundation(line)]%
        {openssl}
\bibfield{author}{\bibinfo{person}{Inc. OpenSSL~Foundation}.}
  \bibinfo{year}{2022 [online]}\natexlab{}.
\newblock \bibinfo{booktitle}{\emph{OpenSSL}}.
\newblock
\urldef\tempurl%
\url{https://www.openssl.org/news/vulnerabilities.html}
\showURL{%
Retrieved November 11, 2022 from \tempurl}


\bibitem[Poyarekar(2021)]%
        {underflow1}
\bibfield{author}{\bibinfo{person}{Siddhesh Poyarekar}.}
  \bibinfo{year}{2021}\natexlab{}.
\newblock \bibinfo{booktitle}{\emph{Bug 28769 (CVE-2021-3999) - Off-by-one
  buffer overflow/underflow in getcwd()}}.
\newblock
\urldef\tempurl%
\url{https://sourceware.org/bugzilla/show_bug.cgi?id=28769}
\showURL{%
\tempurl}


\bibitem[Serebryany(2016)]%
        {DisccusionOfMPX}
\bibfield{author}{\bibinfo{person}{Konstantin Serebryany}.}
  \bibinfo{year}{2016}\natexlab{}.
\newblock \bibinfo{booktitle}{\emph{Discussion of Intel Memory Protection
  Extensions (MPX) and comparison with AddressSanitizer}}.
\newblock
\urldef\tempurl%
\url{https://github.com/google/sanitizers/wiki/AddressSanitizerIntelMemoryProtectionExtensions.}
\showURL{%
\tempurl}


\bibitem[Song et~al\mbox{.}(2019)]%
        {Sanitizer}
\bibfield{author}{\bibinfo{person}{Dokyung Song}, \bibinfo{person}{Julian
  Lettner}, \bibinfo{person}{Prabhu Rajasekaran}, \bibinfo{person}{Yeoul Na},
  \bibinfo{person}{Stijn Volckaert}, \bibinfo{person}{Per Larsen}, {and}
  \bibinfo{person}{Michael Franz}.} \bibinfo{year}{2019}\natexlab{}.
\newblock \showarticletitle{SoK: Sanitizing for Security}. In
  \bibinfo{booktitle}{\emph{2019 IEEE Symposium on Security and Privacy (SP)}}.
  \bibinfo{pages}{1275--1295}.
\newblock
\urldef\tempurl%
\url{https://doi.org/10.1109/SP.2019.00010}
\showDOI{\tempurl}


\bibitem[Travitch(line)]%
        {wllvm}
\bibfield{author}{\bibinfo{person}{Travitch}.} \bibinfo{year}{2022
  [online]}\natexlab{}.
\newblock \bibinfo{booktitle}{\emph{Travitch/whole-program-LLVM: A wrapper
  script to build whole-program LLVM bitcode files}}.
\newblock
\urldef\tempurl%
\url{https://github.com/travitch/whole-program-llvm}
\showURL{%
\tempurl}


\bibitem[Woodruff et~al\mbox{.}(2014)]%
        {CHERI}
\bibfield{author}{\bibinfo{person}{Jonathan Woodruff},
  \bibinfo{person}{Robert~N.M. Watson}, \bibinfo{person}{David Chisnall},
  \bibinfo{person}{Simon~W. Moore}, \bibinfo{person}{Jonathan Anderson},
  \bibinfo{person}{Brooks Davis}, \bibinfo{person}{Ben Laurie},
  \bibinfo{person}{Peter~G. Neumann}, \bibinfo{person}{Robert Norton}, {and}
  \bibinfo{person}{Michael Roe}.} \bibinfo{year}{2014}\natexlab{}.
\newblock \showarticletitle{The CHERI Capability Model: Revisiting RISC in an
  Age of Risk}. In \bibinfo{booktitle}{\emph{Proceeding of the 41st Annual
  International Symposium on Computer Architecuture}} (Minneapolis, Minnesota,
  USA) \emph{(\bibinfo{series}{ISCA '14})}. \bibinfo{publisher}{IEEE Press},
  \bibinfo{pages}{457–468}.
\newblock
\showISBNx{9781479943944}


\bibitem[W\"{u}rthinger et~al\mbox{.}(2007)]%
        {ArrayBondsCheck}
\bibfield{author}{\bibinfo{person}{Thomas W\"{u}rthinger},
  \bibinfo{person}{Christian Wimmer}, {and} \bibinfo{person}{Hanspeter
  M\"{o}ssenb\"{o}ck}.} \bibinfo{year}{2007}\natexlab{}.
\newblock \showarticletitle{Array Bounds Check Elimination for the Java
  HotSpot™ Client Compiler}. In \bibinfo{booktitle}{\emph{Proceedings of the
  5th International Symposium on Principles and Practice of Programming in
  Java}} (Lisboa, Portugal) \emph{(\bibinfo{series}{PPPJ '07})}.
  \bibinfo{publisher}{Association for Computing Machinery},
  \bibinfo{address}{New York, NY, USA}, \bibinfo{pages}{125–133}.
\newblock
\showISBNx{9781595936721}
\urldef\tempurl%
\url{https://doi.org/10.1145/1294325.1294343}
\showDOI{\tempurl}


\bibitem[Xu et~al\mbox{.}(2021)]%
        {In-Fat}
\bibfield{author}{\bibinfo{person}{Shengjie Xu}, \bibinfo{person}{Wei Huang},
  {and} \bibinfo{person}{David Lie}.} \bibinfo{year}{2021}\natexlab{}.
\newblock \showarticletitle{In-Fat Pointer: Hardware-Assisted Tagged-Pointer
  Spatial Memory Safety Defense with Subobject Granularity Protection}
  \emph{(\bibinfo{series}{ASPLOS '21})}. \bibinfo{publisher}{Association for
  Computing Machinery}, \bibinfo{address}{New York, NY, USA},
  \bibinfo{pages}{224–240}.
\newblock
\showISBNx{9781450383172}
\urldef\tempurl%
\url{https://doi.org/10.1145/3445814.3446761}
\showDOI{\tempurl}


\end{thebibliography}

\end{document}